\documentclass[doublecol]{epl2}  
\usepackage{graphicx}
\usepackage{dcolumn}
\usepackage{bm}
\usepackage[dvips]{epsfig}

\newcommand{\ep}{\epsilon}
\newcommand{\kint}{{\cal T}}

\renewcommand{\d}{{\rm d}}
\newcommand{\BEA}{\begin{eqnarray}}
\newcommand{\EEA}{\end{eqnarray}}
\newcommand{\bs}{{\bf s}}
\newcommand{\Pik}{{\cal P}}
\newcommand{\ba}{{\bf a}}
\newcommand{\bbs}{\bar{{\bf s}}}

\title{Kinetics of the helix-coil transition}

\author{Armen E. Allahverdyan, \inst{1} Sasun G. Gevorkian, \inst{1}
Aleksandr Simonian \inst{2} }
\shortauthor{A.E. Allahverdyan, S.G. Gevorkian }

\institute{
\inst{1}
Yerevan Physics Institute, Alikhanian Brothers Street 2, Yerevan 375036, Armenia \\
\inst{2}
Materials Research and Education Center,
275 Wilmore Auburn University, Auburn AL 36849-5341 }

\abstract{ Based on the Zimm-Bragg model we study cooperative helix-coil
transition driven by a finite-speed change of temperature.  There is an
asymmetry between the coil$\to$helix and helix$\to$coil transition: the
latter is displayed already for finite speeds, and takes shorter time
than the former. This hysteresis effect has been observed
experimentally, and it is explained here via quantifying system's
stability in the vicinity of the critical temperature.  A finite-speed
cooling induces a non-equilibrium helical phase with the correlation
length larger than in equilibrium. In this phase the characteristic
length of the coiled domain and the non-equilibrium specific heat can
display an anomalous response to temperature changes. Several pertinent
experimental results on the kinetics helical biopolymers are discussed in detail. }


\begin{document}

\maketitle

Biopolymers carry information which is embedded not only in the linear
sequence of the monomers, but also in the conformational structures
\cite{book}.  These structures are determined by the sequence, but they
also adjust to environmental conditions.  A pertinent example is the
helix-coil transition (HCT), which denotes the disruption of the ordered
conformation|e.g., the $\alpha$-helix of proteins or the triple-helix of
collagen|to form a disordered coil \cite{book}. This
order-disorder transition occurs when the temperature is raised or a
chemical denaturant is added. 

The importance of helices in biopolymers motivated many studies on
thermodynamic and kinetic aspects of HCT 
\cite{book,qian,zimmo,go_go,takano_eq,kin,binder,takano_neq}. 
The basic model in this field was proposed by Zimm and Bragg 
(ZB) and successfully applied for
describing the HCT both in \cite{book,qian,zimmo,go_go,takano_eq} and
out \cite{kin,huang,chay,binder,takano_neq,porsch,jun} of equilibrium. 
The virtue of the model is the simplicity of its ingredients: 
cooperativity and the free-energy preference to form a helix. 
Many studies devoted to the kinetics of the HCT concentrate on the
relaxation from a non-equilibrium state, which is prepared
experimentally via, e.g., laser-induced temperature jump
\cite{takano_neq}. This setup is adequate for small globular proteins
whose relaxation time is short. The situation is different for HCT in
collagen \cite{davis,engel,collagen,miles}, which has very long
relaxation times to equilibrium (hours and days), so that the HCT has to
be probed by necessarily finite-rate temperature changes, and the very
description of the HCT has to be essentially kinetic \cite{miles} (the 
equilibrium HCT still exists at unrealistically long
times \cite{collagen}). Kinetic effects are also encountered in HCT for 
DNA \cite{anshelevich,books} and for crystals
of globular proteins, where each site of the crystal contains one protein \cite{sg}. 
This ordered protein ensemble amplifies memory and hysteresis effects and allows their clear
experimental identification \cite{sg}. 

These experimental set-ups call for a unifying theoretical approach.
Here we study the helix-coil kinetics of ZB model driven by finite-speed
temperature changes. We reproduce and explain several basic experimental
findings, and also predict new effects. Conceptually, there is an even
deeper interest in the kinetics of the basic ZB model, since the
stability of many proteins does have both kinetic and thermodynamic
aspects \cite{kinos}. 

In the ZB model one assigns the spin variable $s_i=1$ ($s_i=-1$) for 
the $i$'th helix (coiled) region of the polymer, and assumes 
the following free energy for the spins \cite{book,qian}
\BEA
\label{kut1}
\label{hamo}
F[\bs]=-J{\sum}_{i=1}^{N-1} s_i s_{i+1}-h(T){\sum}_{i=1}^N s_i,
\EEA
where $\bs\equiv(s_1,.. s_N)$, $N$ is the total number of regions, and
$J>0$ stands for the cooperative interaction \cite{football}. The mechanisms of
cooperativity for the main biopolymers ($\alpha$-helices,
DNA, collagen, {\it etc}) are reviewed in \cite{book,qian,engel}. 

The term $h(T)$ in (\ref{hamo}) is the free energy difference 
of fast atomic variables in region $i$, 
calculated for a fixed value of the slow spin $s_i$ \cite{foot_1}. We thus assume
time-scale separation \cite{binder,takano_neq}: the joint probability $P(\bs,\ba)$ 
of fast ($\ba$) and slow ($\bs$) variables
factorizes as $P({\bf s} )P_{eq}(\ba|\bs)$, where the conditional probability
$P_{eq}(\ba|\bs)$ is always (also for kinetic processes) at equilibrium with
the bath temperature $T$ \cite{foot_1}. The dynamics of the spins 
is then governed by the free energy (\ref{kut1}), where
$h(T)$ favors helix (coil) formation at low
(high) $T$: $h(T_c)=0$ at the equilibrium HCT temperature $T_c$,
while $h(T)>0$ ($h(T)<0$) for $T<T_c$ ($T>T_c$).  Experiments and {\it ab
initio} calculations are consistent with a linear change of $h(T)$ in the
vicinity of $T_c$ ($\alpha>0$ is a constant and $k_B=1$)
\cite{book,qian,zimmo,go_go}:
\BEA
\label{kut2}
\label{kobra10}
h(T)=\alpha(T_c-T).
\EEA 
The interaction strength in the model is characterized by $\sigma=e^{-4J/T_c}$
\cite{qian}.  In the highly-cooperative regime $\sigma\ll 1$ the
equilibrium helix-coil transition resembles a real phase-transition,
which combines the features of first-order (jumping order parameter) and
the second-order (large correlation length) phase transitions
\cite{book,qian,azbel}. 

The ZB model was originally proposed for describing the
$\alpha$-helix-coil transition in polypeptides and proteins
\cite{book,qian}. Later on the equilibrium ZB model was successfully
applied to the duplex-coil transition in DNA \cite{azbel} and to some
aspects of the triplex-coil transition in collagen \cite{engel}. In the
latter two cases the ZB model is regarded as a skeletal model producing
important qualitative conclusions. The purpose of this Letter is to
understand the basic physics of hysteresis and memory effects during the
helix-coil transition in terms of the ZB model. 

Now the system described by (\ref{hamo}) interacts with a bath at
temperature $T$. We assume that the elementary bath-driven process
amounts to local disruption (or creation) of a single helix: $s_j\to
-s_j$ (spin-flipping), and that the dynamics is given by the following
master equation \cite{glauber}:
\BEA
\dot{P}(\bs,t)={\sum}_{j=1}^N\left[
P(\bbs_j,t)w(\bs|\bbs_j)-P(\bs,t)w(\bbs_j|\bs)
\right],
\label{markoo}
\EEA
where $\bbs_j=(s_1,..,s_{j-1},-s_{j},s_{j+1},.., s_N)$, $P(\bs,t)$ is
the time-dependent probability of $\bs=(s_1,..,s_N)$,
$\dot{P}\equiv\partial_tP$, and where the first (second) term in the RHS
of (\ref{markoo}) describes the in-flow (out-flow) of probability to the
configuration $\bs$ due to spin-flipping.  $w(\bs|\bbs_j)$ is the
transition rate $\bbs_j\to\bs$ which is standardly taken in the Glauber
form \cite{glauber}: $w(\bbs_j|\bs)=\frac{\Gamma }{2}[1-s_j\tanh (\beta
\mu_j)]$, where $\Gamma$ is the relaxation frequency, $\beta=1/T$ and
$\mu_j\equiv h+J(s_{j-1}+s_{j+1})$ is the local field acting on $s_j$; see (\ref{kut1}).
Thus the spin-flip is probable if it decreases $F$. This ensures
relaxation to the equilibrium $P_{\rm eq}[\bs]\propto e^{-\beta F[\bs]}$
for a constant $T$ \cite{glauber}.  

Note that the spin-flipping can occur anywhere in the chain. Thus, the
studied kinetics of the ZB model differs from the zipper kinetics
\cite{chay,jun,anshelevich,porsch}, where the disruption of the helix
can occur only at the end-points of the chain. The zipper kinetics is
expected to be valid for relatively short chains, undergoing relaxation
from the completely helical to the completely coiled chains \cite{chay}.
Here we consider long chains. 

Let us introduce the following averaged quantities:
\BEA
m(t)\equiv\langle s_i\rangle_t, \quad
\ep(t)\equiv\langle s_i s_{i+1}\rangle_t, \quad
\ep_2(t)\equiv\langle s_i s_{i+2}\rangle_t,
\label{tartar} 
\EEA
where $\langle\ldots\rangle_t$ is the average over 
$P(\bs,t)$, while $\frac{1+m}{2}$ is the fraction of helical
regions. For $N\gg 1$ (long chain) the boundary effects are neglected,
all spins are equivalent, and (\ref{markoo}, \ref{tartar}) imply
\BEA
\label{kk1}
&&\Gamma^{-1}\,\dot{m}=-(1-a_1)m+\frac{a_0}{2}
+\frac{a_2}{2}\ep_2,\\
\label{kk2}
&&\Gamma^{-1}\,\dot{\epsilon}=-2\epsilon+a_1+(a_0+a_2)m
+a_1\ep_2,\\
\label{kk33}
&&a_{1} =\kappa_+ - \kappa_-,\quad
a_{0,2} =\pm \tanh (\beta h)
+\kappa_+ + \kappa_- ,
\EEA
where $\kappa_\pm= \frac{1}{2}\tanh (\beta h\pm 2\beta J)$.
Eqs.~(\ref{kk1}, \ref{kk2}) are first two equations of
the infinite hierarchy of moment equations. For $h=m=0$ this
hierarchy is exactly solvable \cite{glauber}.
For $h\not =0$, there is no exact solution, and one has to rely 
on approximations \cite{kin,binder}. 

\begin{figure}
\includegraphics[width=6.5cm]{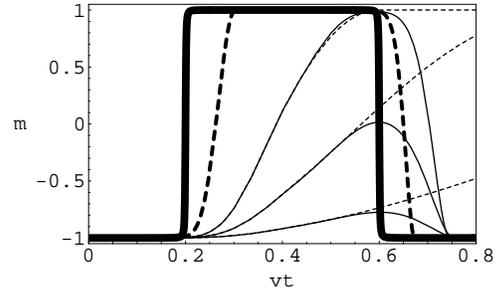}
\caption{  
The order parameter $m$ versus dimensionless time $vt$ 
under cooling-reheating with temperature speed $v$ and parameters
(\ref{mara}), except that $\sigma=1.2\times 10^{-6}$.  
Cooling starts at $vt=0$ and changes to reheating
at $vt=0.4$. The equilibrium $T_c$ is crossed for $vt=0.2$ and $t=0.6$.
Thick curve: equilibrium $m_{eq}$, which changes between $m_{eq}=-1$ (coil)
and $m_{eq}=1$ (helix).  Thick dashed curve:
$v/\Gamma=2\times 10^{-6}$.  Normal curves (from bottom to top):
$v/\Gamma=5\times 10^{-5},\, 2\times 10^{-5},\, 10^{-5}$. The dashed
counterpart of each normal curve refers to cooling from $vt=0$ till
$vt=0.4$, and then holding $T$ constant. 
}
\label{ff_0}
\end{figure}

{\it The spin-temperature anzatz} amounts to assuming that the probability
$P(\bs,t)$ has a locally Gibbsian form with two time-dependent parameters
$\beta_1$ and $\beta_2$ \cite{goldman,berim}:
\BEA
\label{klamo}
P(\bs,t)\propto \exp[\,\beta_1(t)J{\sum}_i s_i s_{i+1}+\beta_2(t)h{\sum}_is_i \,].
\EEA
This amounts to expressing the term $\ep_2$ in (\ref{kk1}, \ref{kk2}) 
via $\ep$ and $m$ by means of equilibrium formulas \cite{berim}
\BEA
\label{kk3}
\epsilon_k=m^2+(1-m^2)^{1-k}(\epsilon-m^2)^k, \qquad k=2,3...
\EEA
Thus we assume that the higher-order moments
$\ep_{k\geq 2}$ relax to the local equilibrium (\ref{kk3}), before
$m(t)$ and $\ep(t)$ relax to equilibrium \cite{goldman,berim,huang}.  Eqs.~(\ref{kk1}, \ref{kk2})
with $\ep_2$ given by (\ref{kk3}) is now a closed pair of equations that
reproduces exactly the equilibrium limit.
Eqs.~(\ref{kk1}--\ref{kk33}, \ref{kk3}) are consistent with the above-mentioned
exact solution for $h=m=0$, since they correctly predict the diverging
relaxation time for $\beta J\gg 1$, as well as the transition from the
exponential to a power-law decay \cite{berim}.  The reliability of
the spin-temperature anzatz is confirmed by its applications
in NMR/ESR physics \cite{goldman,berim}. For the derivations of 
(\ref{klamo}) via the projection-operator method see \cite{rau}.

\begin{figure}
\includegraphics[width=6.5cm]{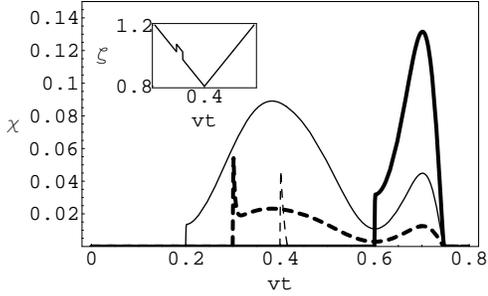}
\caption{Susceptibility $\chi(t,t')$ versus $vt\equiv \tau$ 
under cooling-reheating with parameters (\ref{mara}), except that
$\sigma=1.2\times 10^{-6}$ and $v/\Gamma= 10^{-5}$.
The temperature perturbation 
$\varepsilon T_c\theta[\tau-\tau'+\frac{\varepsilon}{2}]
\theta[\tau'+\frac{\varepsilon}{2}-\tau]$
has magnitude $\varepsilon=0.001$ and is centered at $\tau$. Normal curve: $\tau'=0.2$; thick-dashed: $\tau'=0.3$;
dashed: $\tau'=0.41$; thick: $\tau=0.6$. {\bf Insert}: the temperature profile $\zeta(vt)=T(vt)/T_c$ 
perturbed at $vt=0.2$.
}
\label{ff_01}
\end{figure}

The equilibrium $m$ is obtained from putting to zero the LHS of
(\ref{kk1}, \ref{kk2}) with $\ep_2$ given by (\ref{kk3}): $m_{eq}=\sinh
(\beta h)[\sinh^2( \beta h)+e^{-4 \beta J}]^{-1/2}$.  In the cooperative
regime $\sigma=e^{-4 \beta_c J}\ll 1$, $m_{eq}$ shows a sharp transition
from $m_{eq}=-1$ (coil) to $m_{eq}=1$ (helix) \cite{book}; see Fig.
\ref{ff_0}. 

Note that for $h=0$ (right at $T=T_c$) (\ref{kk1}, \ref{kk33}) predict
that the relaxation time of $m(t)$ is
$\frac{1}{\Gamma(1-a_1)}=\frac{1+\sigma}{2\sigma\Gamma}$.  This agrees
with the known result \cite{kin} and shows that the relaxation time is
larger for more cooperative transitions and for lower temperatures.
These aspects were numerously confirmed in simulations \cite{jun}.  

Now assume that the {\it bath temperature} $T(t)/T_c=\zeta_0-vt$ decreases with 
speed $v$ from a higher temperature $T_c\zeta_0>T_c$ (at which the
system was equilibrated) to a lower temperature $T_c\zeta_1<T_c$. Then
$T(t)$ increases back to $T_c\zeta_0$ with the same speed (reheating):
$T(t)/T_c=\zeta_1+vt$.  For identification of asymmetries in the
system response, the reheating temperature profile is taken to be the
mirror reflection of the cooling profile. 

For concreteness the numerical solutions of (\ref{kk1}, \ref{kk2}) are
displayed for the dimensionless parameters [see (\ref{kobra10})]: 
\BEA
\label{mara}
\zeta_0=1.2,~~
\zeta_1=0.8,~~ \alpha=0.75,~~ \sigma=3.34\times
10^{-4}, 
\EEA
and various values of the dimensionless cooling-reheating speed
$v/\Gamma$. The values of $\alpha$ and $\sigma$ correspond to the
helix-coil transition in poly-$\gamma$-benzyl-glutamate \cite{zimmo}.
They are typical for other cooperative helix-coil transitions
\cite{go_go}. 

{\it The order parameter} $m$ defines the helicity fraction
$\frac{1+m}{2}$.  Fig.~\ref{ff_0} displays the non-equilibrium $m$ versus
the dimensionless time $vt$, as obtained from solving numerically
(\ref{kk1}, \ref{kk2}, \ref{kk3}) with the time-dependent temperature
$T(t)$. Fig.~\ref{ff_0} shows that for a small (but finite) speed
$v/\Gamma$ the transition helix$\to$coil during the reheating is more
visible and takes shorter time than the reverse transition
coil$\to$helix during the cooling.  The same conclusion (not displayed)
holds when doing heating and then recooling. 
The symmetry between helix$\to$coil and
coil$\to$helix is recovered in the equilibrium 
limit $\frac{v}{2\sigma\Gamma}\ll 1$.  
The asymmetry also disappears in the weakly-cooperative case
$\sigma\simeq 1$.  This asymmetry is an example of hysteresis and it was
observed experimentally for highly-cooperative HCT in collagen
\cite{davis,engel,collagen,miles}, crystalline proteins \cite{sg} and DNA \cite{books}.  

To gain a deeper understanding of the hysteresis, let us study in more detail
the system's memory. Compare the
cooling-reheating behavior of $m(t)$ with the situation, where the
temperature decreases till the lowest point and is then held constant;
see Fig.~\ref{ff_0}. This cooling-holding scenario produces curves which
are almost identical to the cooling-reheating curves, except at the vicinity of
the equilibrium reheating transition, i.e., $m(t)$ does not react on the reheating
before the sign of $h$ changes. Let us 
quantify the memory via susceptibility 
\BEA
\chi(t,t')={\rm
lim}_{\varepsilon\to 0}[m(t)-\widetilde{m}(t)]/\varepsilon.
\EEA
Here $\widetilde{m}(t)$ is obtained under the same cooling-reheating
temperature setup, but $\varepsilon$-perturbed at $t'$:
\BEA
\widetilde{T}(t)/T_c=\zeta_0-vt+\varepsilon
\theta[vt+\delta-vt']\theta[vt'+\delta-vt] , 
\EEA
where $\theta[t]$ is the
step function, and where the perturbation duration $2\delta$ is small
but finite; see Fig.~\ref{ff_01}. The perturbation is designed such that
at equilibrium, where $m(t)$ is a function of the time-dependent
temperature $T(t)$, $\chi(t,t')$ is non-zero only for $v|t-t'|<\delta$.
In the regime where the above asymmetry is present, there are basically
three scenarios for the behavior of $\chi(t,t')$; see Fig.~\ref{ff_01}.
{\it i)} A perturbation introduced during the cooling in the vicinity of
$T_c$ is {\it memoryzed and amplified}.
This memory need not be monotonic: it revives once $T(t)$
crosses $T_c$ during the reheating; see Fig.~\ref{ff_01}. {\it ii)} The
same perturbation introduced during the reheating in the vicinity of
$T_c$ creates a stronger immediate response, but a weaker memory, as
compared to the previous case.  {\it iii)} Outside of the vicinity of
$T_c$ the response resembles that in equilibrium: $\chi(t,t')$ is
maximal for $t\approx t'$ and quickly decays for $|t-t'|>0$; see
Fig.~\ref{ff_01}. We thus see how 
the hysteresis emerges from the unstabilities at $T(t)\simeq T_c$ during the cooling
and reheating. 

\begin{figure}[ht]
\vspace{0.1cm}
\includegraphics[width=6.5cm]{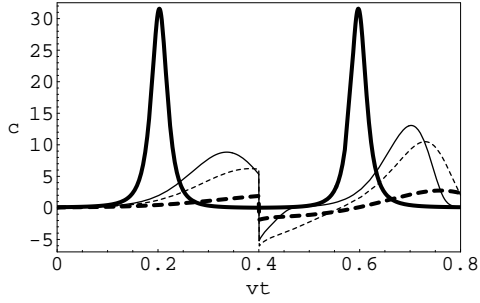}
\caption{Specific heat $c$
versus the dimensionless time $vt$  
under cooling-reheating with parameters (\ref{mara}).
Normal curve: $v/\Gamma=\frac{1}{5}\times 10^{-3}$.  Dashed curve: $v/\Gamma=\frac{1}{3}\times 10^{-3}$.
Dashed-thick curve: $v/\Gamma=10^{-3}$. 
Thick curve: equilibrium $c_{eq}$.}
\label{ff_3}
\end{figure}

\begin{figure}
\includegraphics[width=6.5cm]{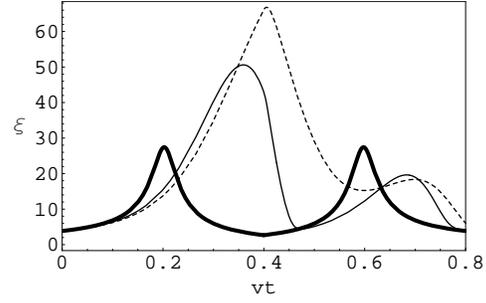}
\caption{
Correlation length $\xi$ 
versus $vt$  
under cooling-reheating with parameters (\ref{mara}).
Normal curve: $v/\Gamma=\frac{1}{5}\times 10^{-3}$.  Dashed curve: $v/\Gamma=\frac{1}{3}\times 10^{-3}$.
Thick curve: equilibrium
$\xi_{eq}$.}
\label{ff_4}
\end{figure}

{\it Specific heat} is a well-known indicator of the helix-coil
transitions, which are observed via calorimetric methods \cite{book}.
Recalling the discussion before (\ref{kut2}) and definitions
(\ref{kk33}), one can see that the energy of the spins is $N u(t)$
\cite{foot_1}, with
\BEA
\label{krot}
u(t)=-J\ep(t)-\partial_\beta [\beta h(T)]\,m(t).  
\EEA
The origin of this formula should be clear from 
(\ref{hamo}, \ref{tartar}). Note that the term $-\partial_\beta [\beta h(T)]$ is 
similar to the known equilibrium formula $E=-\partial_\beta[\beta F]$
relating energy $E$ to the free energy $F$. The specific heat $c(t)$ is
the response of $u(t)$ to the temperature change [$\dot T\equiv \d T/\d t$]
\BEA
\label{boro}
c(t)\equiv{\dot{u}(t)}/{\dot{T}(t)}
=-\left[
J\dot{\ep}+\alpha T_c\, \dot{m}
\right]/{\dot{T}}.
\EEA
The equilibrium specific heat $c_{eq}$ is always positive and shows two
sharp and symmetric peaks at $T=T_c$; see Fig.~\ref{ff_3}. For the
non-equilibrium specific heat $c$ we see again the asymmetry between
helix$\to$coil and coil$\to$helix transitions: the peak of $c$ during
cooling is either absent or less visible than the one during the
reheating.  Now $c(t)$ can be negative, i.e., the internal energy can
decrease upon reheating; see Fig.~\ref{ff_3}.  This is partially related
to the response of $m(t)$: Fig.~\ref{ff_0} shows that when cooling
changes to reheating, $m(t)$ continues to increase due to its memory.
In contrast, $m_{eq}$ decreases under reheating. The negative part of
$c$ is most pronounced for a finite cooling-reheating speed $v$; see
Fig.~\ref{ff_3}.  Note that a negative specific heat is met in glasses,
within a different scenario that is also related to large relaxation
times \cite{bis}. 

The kinetic transition temperature $\kint$ can be related to the peak of
the specific heat \cite{collagen,miles}; see Fig.~\ref{ff_3}. For not
very small $v$, $\kint$ is approximately a linear function of
$x\equiv\ln\left[\frac{v}{2\sigma\Gamma}\right]$; thus $\kint$ is not
susceptible to moderate changes in the cooling-reheating speed $v$.  The
same scaling of the kinetic transition temperature was seen
experimentally for the helix-coil transition of collagen
\cite{collagen}. For the parameters of Fig.~\ref{ff_3} we obtained
$(\kint_{\bf \rm C,R}-T_c)/T_c= a_{\rm C,R} x+ b_{\rm C,R}$ for $x\in
[-2,-3.5]$.  Here the lower indices ${\rm C}$ and ${\rm R}$ refer to the
cooling and reheating, respectively, while $a_{\rm R}=0.09$, $a_{\rm
C}=-0.12$, $b_{\rm R}=0.32$, and $b_{\rm C}=-0.41$. We see that
$\kint_{\rm R}>T_c>\kint_{\rm C}$. 

{\it The correlation function} $g(k,t)=\langle s_{i}s_{i+k}\rangle_t-\langle
s_{i}\rangle_t\langle s_{i+k}\rangle_t$ describes the spatial structure 
of fluctuations. 
Eq.~(\ref{kk3}) implies $g(k,t)=[1-m^2(t)]\, e^{-k/\xi(t)}$, where
\BEA
\label{corro}
\xi(t)=\left(\ln\left[\frac{1-m^2(t)}{\epsilon(t)-m^2(t)}\right]\right)^{-1},
\EEA
is the correlation length, or the cooperative unit length, which plays
an important role in describing the cooperativity of the polymer structure \cite{book}. The
equilibrium $\xi_{eq}$, obtained from (\ref{corro}) by $m(t)\to m_{eq}$
and $\ep(t)\to \ep_{eq}$, displays two sharp peaks at $T=T_c$; see
Fig.~\ref{ff_4}. It directly relates to the cooperativity parameter:
$\xi_{eq}(T=T_c)=\frac{1}{2\sqrt{\sigma}}$ \cite{book}.  
$\xi_{eq}$ is small both below and above $T_c$, 
since there are no much equilibrium fluctuations 
there. 

The non-equilibrium behavior of $\xi(t)$ is different: for finite speeds
$\xi(t)$ increases during cooling, and its maximum is reached for the
lowest $T$; see Fig.~\ref{ff_4}. $\xi(t)$ is maximal at a certain finite
speed of the cooling-reheating. Around its maximum it is {\it larger}
than $\xi_{eq}(T=T_c)$. Thus the equilibrium relation between step-like
change of the order parameter $m$ and the correlation length is broken
in kinetics: now $m(t)$ does not show transition during cooling, but
$\xi(t)$ is large.  The reason for a large $\xi(t)$ is that the
spin-spin interaction energy $\ep(t)$ is close to $1$ for both $T<T_c$
and $T>T_c$, while $m(t)$ is far from $\pm 1$; see (\ref{corro}) and
Fig.~\ref{ff_0}.  Thus there are many helical and coiled domains, whose
total spin should sum to zero implying long correlations. Yet another
interpretation of a large $\xi(t)$ is that due to memory the system does
not see the sign change of $h(T)$, but it sees the lowering of $T(t)$,
which naturally increases its correlation length $\xi(t)$. Thus the
finite-rate cooling plays a selective role suppressing one mechanism
and activating another. 
In the non-cooperative case $\sigma\simeq 1$, 
$\xi(t)$ follows to the shape of $\xi_{eq}$, and
$\xi(t)<\xi_{eq}$ for all temperatures.

\begin{figure}
\includegraphics[width=6cm]{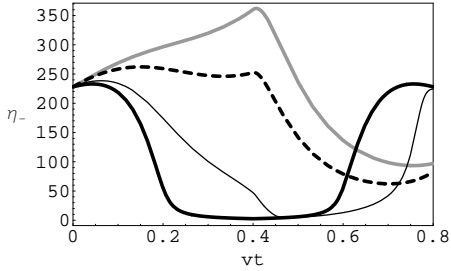}
\caption{The coiled domain length $\eta_-$ 
versus $vt$  
under cooling-reheating with parameters (\ref{mara}).
Gray curve: $v/\Gamma=\frac{1}{5}\times 10^{-2}$.  Thick-ashed curve: $v/\Gamma= 10^{-3}$.
Normal curve: $v/\Gamma= \frac{1}{5} \times 10^{-3}$.
Thick curve: equilibrium
$\eta_{-,\,eq}$.}
\label{ff_6}
\end{figure}
\begin{figure}
\includegraphics[width=6cm]{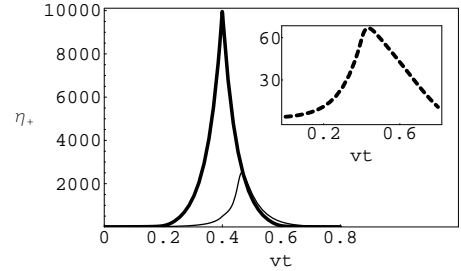}
\vspace{0.1cm}
\caption{
The helical domain length $\eta_+$ 
versus the dimensionless time $vt$  
under cooling-reheating with parameters (\ref{mara}).
Normal curve: $v/\Gamma= \frac{1}{5} \times 10^{-3}$.
Thick curve: equilibrium
$\eta_{+,\,eq}$. {\bf Insert}: the same with
$v/\Gamma=10^{-3}$.  
}
\label{ff_7}
\end{figure}

{\it Domain lengths.} Generally, the state of the studied linear polymer 
is inhomogeneous and consists of helical and coiled domains. To characterize 
the domain lengths, we consider the probability 
$\Pik_{\pm}(k)=\left\langle
{\prod}_{j=1}^k\frac{1\pm s_j}{2}\right\rangle_t$
of having a helical ($\Pik_+$) or coiled ($\Pik_-$) domain of length $k$.  The transfer matrix
treatment of (\ref{klamo}) leads to $\Pik_{\pm}(k)= \gamma_\pm
e^{-k/\eta_{\pm}}$, where $\gamma_\pm$ and $\eta_\pm$ do not depend on
$k$. Thus, $e^{-1/\eta_{\pm}}=\Pik_{\pm}(2)/\Pik_{\pm}(1)$, or
\BEA
\eta_{\pm}(t)= \left(\ln\left[\frac{2\pm 2m(t)}{1+\epsilon(t)\pm
2m(t)}\right]\right)^{-1}, 
\EEA
where $\eta_+$ and $\eta_-$ are, respectively, the
characteristics length of the helical and coiled domains.  Naturally, the
equilibrium $\eta_{- \, {eq}}$ is large in the coiled phase and
decreases under cooling becoming small in the helical
phase; see Fig. \ref{ff_6}. Likewise, $\eta_{+ \, {eq}}$ is large
(small) in the helical (coiled) phase. Note that $\eta_{+ \, {eq}}(T_c)=
\eta_{-\, {eq}}(T_c)$. In contrast to the equilibrium $\eta_{-\,eq}$,
we see in Fig.~\ref{ff_6} that
$\eta_{-}(t)$ can {\it increase} in time, if the cooling is not very slow. 
It decreases once the reheating starts.  Thus
$\eta_-$ reacts stronger on the decrease of temperature, than on
changing the sign of $h$. Since during the cooling the order parameter
$m$ increases, we see that the number of helical segments increases, but
the typical coiled domain gets larger.  Fig.  \ref{ff_6} also shows that
$\eta_{-}$ can behave non-monotonically with time.  For very small
speeds, $\eta_{-}$ reproduces (with a delay) the shape of $\eta_{-\, eq}$; 
see Fig.~\ref{ff_6}.  The behavior of $\eta_{+}(t)$ under cooling-reheating is less
interesting: it follows, in a delayed and weakened form, the shape of
$\eta_{+\, eq}$; see Fig. \ref{ff_7}.  (The same conclusion holds for
$\eta_{+}$ during the heating from $T<T_c$ and then recooling.) Note
that for finite speeds we can have $\eta_->\eta_+$ at low temperatures;
see Figs. \ref{ff_6}, \ref{ff_7}. 

{\it Relations with experiments.} We now discuss several experimental
results on helical biopolymers that demonstrate clear signs of
irreversibility and hysteresis. 

Admittedly, many experiments in polypeptides and proteins do not show
visible signs of memory and hysteresis (unless caused by irreversible
aggregation), mainly because the experimental temperature changes are
too slow compared to relevant relaxation times \cite{kinos}. 

\begin{figure}[ht]
\includegraphics[width=6.2cm]{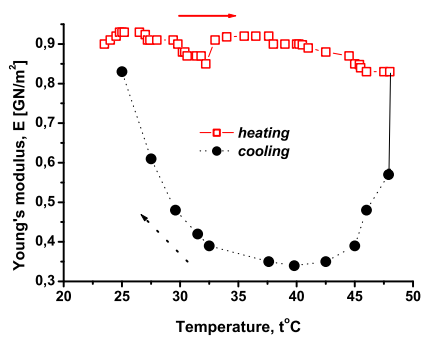}
\caption{
The Young modulus versus temperature for a rhombo-hedric crystall
(direction [z]) each site of which contains Alcohol Dehydrogenase
protein. Molecular weight of this protein is 80 kDa.  The relative
humidity is 97 \% at 25$^\circ$ C. The water content at this temperature
is 0.37 g of water per 1 g of dry protein. The speed of
heating/recooling is 0.1$^\circ$ C per minute. 
}
\label{o1}
\end{figure}

However, there are susceptible experiments on crystals of globular
proteins, which do see memory effects during various conformational
changes \cite{sg}.  Here each site of the crystalline structure contains
one protein.  Memory effects are amplified by this ordered structure and
are visible in experiments \cite{sg}. Along these lines, we present here
several new experimental results. Note that irreversibilities and
hysteresis effects here are not associated with the crystalline
structure {\it per se} \cite{sg}; they are related to helix-coil
features proteins, while the crystalline structure only serves for
amplifying these effects.

Fig.~\ref{o1} shows the denaturation
process for crystalline protein Alcohol Dehydrogenase. The process is
monitored via the change of the Young's modulus with temperature. Recall
that the Young's modulus is defined as the ratio of the applied stress
[pressure] over the induced strain. The Young's modulus serves as an
indicator of structural transitions \cite{Mor,Gev}, since in the
denaturated state it is smaller than in the native state.  The Young's
modulus of an Alcohol Dehydrogenase sample was measured via analyzing
the electrically excited transverse resonance vibrations of the sample,
which is cantilevered from one edge (another edge is free)
\cite{Mor,Gev}. 
The denaturation temperature of Alcohol Dehydrogenase ($\approx
45.5^\circ$ C) is identified via the sudden jump of the Young modulus;
see Fig.~\ref{o1}. This agrees with the denaturation temperature
obtained via calorimetric methods \cite{Privalov}. 

Fig.~\ref{o1} shows that once the heating is substituted by re-cooling in the
vicinity of the critical temperature, the system follows a different
path (hysteresis), although the heating-re-cooling speed was rather
small. Moreover, even though the heating has been changed to
re-cooling, the Young modulus keeps on decreasing till 40$^\circ$ C due
to the memory on the previous heating stage; see Fig.~\ref{o1}. 
These effects agree qualitatively with the theoretical results found above
via ZB model. 

\begin{figure}[ht]
\includegraphics[width=6.2cm]{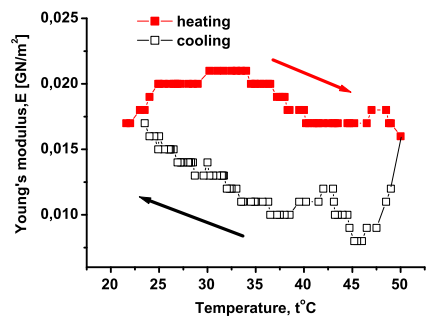}
\caption{
The Young modulus versus temperature for an amorphous DNA film. 
Each DNA macromolecules (taken from a sturgeon mail) 
weights 1 000 kDa. The relative humidity is 95 \% at 25$^\circ$ C. 
The water content at this temperature is 0.42 g of water per 1 g 
of dry DNA. The speed of heating/recooling is 0.1$^\circ$ C per minute.
}
\label{o2}
\end{figure}

In contrast to polypeptides and proteins, there are established
experimental results concerning the hysteresis and memory effects for
DNA \cite{anshelevich,books} and collagen \cite{davis,engel,collagen,miles}. In
both these biopolymers the helical state (duplex for DNA and triplex for
collagen) is stabilized by inter-molecular (i.e., inter-strand)
interactions. 

Fig.~\ref{o2} displays the experimental denaturation of an amorphous DNA
film.  The same effects of hysteresis and memory are present here. 
For other experimental indications of memory and
hysteresis effects during DNA denaturations see Refs.~\cite{books}.
Ref.~\cite{azbel} critically assesses the applicability of the
Zimm-Bragg model to the helix-coil transition in DNA, and finds that
many experimental aspects of this complex phenomenon are adequately
reflected in the equilibrium Zimm-Bragg model. 

Ref.~\cite{engel} investigates the equilibrium Zimm-Bragg model in the
context of the helix-coil transition in collagen III. For this
biopolymer the end-points of the three strands are held together by
disulfide bonds, which precludes mismatches during the renaturation and
makes possible the application of the Zimm-Bragg model. Indeed, it was
found that although experimentally the helix-coil transition in collagen
III (as well as in collagen I) is always kinetic|the proper equilibrium
regime is hardly reached within the experimental observation time|some
important aspects of the phenomenon can be described within the
equilibrium Zimm-Bragg model in quantitative agreement with experiments
\cite{engel}.  We thus expect that the kinetics of the Zimm-Bragg model
can describe the qualitative aspects of memory and hysteresis. 


{\it In sum}, based on the Zimm-Bragg model we studied kinetics of the
helix-coil transition driven by a finite-speed temperature change.  We
reproduced well-known experimental results on the hysteresis during the
kinetic transition and explained it by quantifying the process memory.
We also predicted new scenarios of kinetic helix-coil transition related
to {\it i)} negative non-equilibrium specific heat accompanying the
hysteresis; {\it ii)} correlation length becoming larger than in
equilibrium.

There is an increasing evidence that the characteristics of many
important biopolymers is controlled by both kinetic and thermodynamic
factors \cite{kinos}. For instance, the helix-coil transitions in
collagen are normally kinetic, because the equilibrium is not reached
within reasonable times. Moreover, the kinetic helix-coil transition
temperature of collagen for various organisms is close to their
physiological temperature \cite{collagen}, since this kinetic transition
plays a role in achieving the flexibility of the collagen fiber
\cite{collagen}. Since our results indicate on new scenarios of kinetic
helix-coil transitions in the basic Zimm-Bragg model, they can be
relevant for understanding the interplay between the kinetics and
thermodynamics in biopolymers. 

A.A. thanks Y. Mamasakhlisov for discussions. The work was supported by
Volkswagenstiftung, ANSEF and SCS of Armenia (grant 08-0166).


\begin{thebibliography}{0}
\bibitem{football}Note that the ZB model has several different appearances. All of them
are equivalent to (\ref{hamo}) for a long chain.
\end{thebibliography}
\end{document}